\documentclass[aps,prb,twocolumn,superscriptaddress,showpacs]{revtex4-1}
\usepackage{graphicx}

\begin{document}

\title{Dynamical properties of ordered Fe-Pt alloys}

\author{M. Sternik}
\affiliation{Institute of Nuclear Physics, Polish Academy of
Sciences, Radzikowskiego 152, 31-342 Krak\'{o}w, Poland}

\author{S. Couet}
\affiliation{Instituut voor Kern- en Stralingsfysica, KU Leuven, Celestijnenlaan 200D, B-3001 Leuven, Belgium}

\author{J. \L{}a\.{z}ewski}
\affiliation{Institute of Nuclear Physics, Polish Academy of
Sciences, Radzikowskiego 152, 31-342 Krak\'{o}w, Poland}

\author{P. T. Jochym}
\affiliation{Institute of Nuclear Physics, Polish Academy of
Sciences, Radzikowskiego 152, 31-342 Krak\'{o}w, Poland}

\author{K. Parlinski}
\affiliation{Institute of Nuclear Physics, Polish Academy of
Sciences, Radzikowskiego 152, 31-342 Krak\'{o}w, Poland}

\author{A. Vantomme}
\affiliation{Instituut voor Kern- en Stralingsfysica, KU Leuven, Celestijnenlaan 200D, B-3001 Leuven, Belgium}

\author{K. Temst}
\affiliation{Instituut voor Kern- en Stralingsfysica, KU Leuven, Celestijnenlaan 200D, B-3001 Leuven, Belgium}

\author{P. Piekarz}
\affiliation{Institute of Nuclear Physics, Polish Academy of
Sciences, Radzikowskiego 152, 31-342 Krak\'{o}w, Poland}

\date{\today}

\begin{abstract}
The structure, magnetic properties, and lattice dynamics of ordered Fe-Pt alloys with three stoichiometric compositions, Fe$_3$Pt, FePt and FePt$_3$,  have been investigated using the density functional theory. Additionally, the existing experimental data have been complemented by new measurements of the Fe projected phonon density of states performed for the Fe$_3$Pt and FePt$_3$ thin films using the nuclear inelastic scattering technique. The calculated phonon dispersion relations and phonon density of states have been compared with the experimental data.  The dispersion curves are very well reproduced by the calculations, although, the softening of the transversal acoustic mode TA$_1$ leads to some discrepancy between the theory and experiment in Fe$_3$Pt. A very goood agreement between the measured spectra and calculations performed for the tetragonal structure derived from the soft mode may signal that the tetragonal phase with the space group $P4/mbm$ plays an important role in the martensitic transformation observed in Fe$_3$Pt. For FePt$_3$, the antiferromagnetic order appearing with decreasing temperature has been also investigated. The studies showed that the phonon density of states of FePt$_3$ very weakly depends on the magnetic configuration. 

\end{abstract}	

\pacs{}

\keywords{}

\maketitle

\section{Introduction}

The Fe-Pt alloys constitute a very important class of materials because of their various interesting physical properties and high application potential. Crystallographically, they can be either chemically disordered fcc or ordered fcc/fct structures.  According to the Fe-Pt phase diagram \cite{ph_diag}, at high temperatures an fcc solid solution of the components is observed. With decreasing temperature this disordered A1-type structure exhibits order-disorder transformations leading to three different ordered phases with structures and physical properties depending on the chemical composition.  

At temperatures below 1570~K, the tetragonal L1$_0$ structure is formed for the almost equiatomic concentration region from approximately 35 to 55 atomic percent of Pt. In Fe-Pt alloys with lower or larger Pt concentration, the formation of the stable cubic L1$_2$ structures, FePt$_3$ and Fe$_3$Pt, is observed at temperatures below 1120~K and 1620~K, respectively. Moreover, in the region below 670~K the Fe-rich alloys show the anomalously low thermal expansion coefficient (Invar effect)\cite{invar} and undergo a martensitic transformation.\cite{martensite} The martensite struture depends on the Fe concentration and the chemical order.  In the ordered state, the L1$_2$ Fe$_3$Pt phase is stable down to approximately 60 K, whereas the disordered Fe$_3$Pt starts to transform to a bcc martensite already at room temperature. As a precursor of the distorted phase, the softening of the transverse acoustic phonons is observed in the [110] direction, \cite{tajima} however, the role of the soft mode in the mechanism of the structural transition is not fully explained. 

The ordered equiatomic FePt alloy with the L1$_0$ structure is composed of alternately stacked layers of Fe and Pt atoms along c-axis (AuCu-type structure with the $P4/mmm$ space group).  This structure exhibits ferromagnetic (FM) order below T$_c$ = 750 K.  The FePt alloys with stoichiometry around 1:3 can form a cubic phase Fe$_3$Pt or FePt$_3$ (with the  $Pm\bar3m$ symmetry). In Fe$_3$Pt (or FePt$_3$), the Fe (Pt) atoms occupy the cube corners and the Pt (Fe) atoms occupy the face-center positions.

In Fe-Pt alloys, both Fe and Pt atoms carry magnetic moments, however, the Fe moments are significantly larger. The orientation of local magnetic moments  depends on the Fe and Pt concentrations, arrangement of alloy components and temperature.  The disordered A1 phase of FePt and FePt$_3$ is ferromagnetically ordered in contrast to the Fe$_3$Pt disordered alloy that is paramagnetic.  Differently, the ordered FePt$_3$ alloy is paramagnetic while FePt and Fe$_3$Pt are ferromagnetic.  Moreover, below 170 K, FePt$_3$ is antiferromagnetic (AFM-I phase) with wavevector $q_{1} =  \frac{2 \pi}{a} (\frac{1}{2},\frac{1}{2},0)$. \cite{maat_AFM_FePt$_3$} Subsequently, below 80 K it undergoes a second phase transition into an AFM-II phase with wavevector $q_{2} =  \frac{2 \pi}{a} (\frac{1}{2},0,0)$. \cite{maat_AFM_FePt$_3$} It has been discussed, \cite{maat_AFM_FePt$_3$} that the latter AFM state is metastable and is probably induced by antiphase boundaries or small compositional variations. Such a wide variation of magnetic structures in the FePt alloys is evidently a consequence of different atomic configurations around Fe atoms, which in turn, have a considerable effect on the electronic structure of these alloys.

Differences in crystal structures and chemical compositions influence also lattice dynamics of Fe-Pt alloys. 
Experimentally, the knowledge on phonon spectra for Fe-Pt alloys comes from  fitting the phonon dispersion relations obtained from the inelastic neutron scattering (INS) \cite{exp_FePt,exp_FePt3,exp_Fe3Pt,kastner1,kastner2} to the Born-von Karman model and from nuclear inelastic scattering (NIS) measurements \cite{wiele,fultz,Pdos_Fe3Pt_nano,tamada, couet}. The latter technique provides information on the Fe contribution to the phonon density of states (PDOS), and
it has been applied to ordered and disorder polycrystalline Fe-Pt alloys, \cite{wiele,fultz} thin films, \cite{couet} as well as to nanoclusters \cite{Pdos_Fe3Pt_nano,tamada}.
The theoretical studies based on {\it ab initio} methods enabled calculation of the phonon dispersion relations of Fe/Pt multilayers \cite{ptj}, 
estimation of the magnetic anisotropy of Fe/Pt(001) and Pt/Fe/Pt(001) systems \cite{anisotropy_FePt}, 
identification of the structural and magnetic phases in FePt surface alloys \cite{magnetism_FePt}, 
determination of the partial Fe PDOS in ordered FePt \cite{couet} and investigation of the soft mode behavior in Fe$_3$Pt. \cite{gruner}  
For FePt, the first-principles results were compared with the partial Fe PDOS measured by the NIS method. \cite{tamada,couet}
Nevertheless, no direct comparison between the calculated and measured dispersion curves has been performed for any of the Fe-Pt alloys.

In this paper, we present the results of {\it ab initio} studies on the structural, magnetic, and dynamical properties for three ordered alloys: FePt$_3$, FePt, and Fe$_3$Pt.
The obtained results are confronted with previously measured phonon dispersion curves and PDOSs
as well as new NIS measurements performed on Fe$_3$Pt and FePt$_3$ thin films.  
Using the same formalism and software for all these systems allows us to make detailed comparison and discuss the differences
in lattice dynamics of different Fe-Pt alloys that otherwise would not be possible. 
In particular, we are studying the influence of local atomic configurations and magnetic order on phonon spectra.                    
For Fe$_3$Pt, we analyse the low-symmetry phase obtained from the soft mode, and compare the calculated phonon spectra with the low-temperature measurements. 

The paper is organized as follows. In Section \ref{sec2}, the structure, magnetic ordering, and dynamical properties of room temperature phases of ordered Fe-Pt alloys are analysed and compared with the experimental data. In Section \ref{sec3}, the detailed analysis of the anisotropy of phonon density of states and the low-temperature phases of FePt$_3$ and Fe$_3$Pt is presented. Section \ref{sec4} concludes the results. 

\section{Calculations vs measurements\label{sec2}}

\subsection{Structure and magnetic ordering\label{subsec1}}

Three ordered phases of Fe-Pt alloys were modelled by imposing the symmetry restrictions of the $P4/mmm$  (L1$_0$) and $Pm\bar3m$ (L1$_2$) space groups on the crystal structure. Some results for the L1$_0$ equiatomic structure have already been presented in our previous paper. \cite{couet} The calculations of cubic phases have been performed using the same technique.  Structure optimization was achieved using the VASP package.\cite{vasp}
The spin-polarized density functional theory (DFT) calculations were carried out within the generalized gradient approximation using the Perdew, Burke, and Ernzerhof (PBE) functional.\cite{Per96} 
The wave functions were sampled according to Monkhorst--Pack scheme with a {\bf k}--point mesh of (4,4,4). 
The structural calculations were performed on a $2\times 2\times 2$ supercell (containing 32 atoms) with periodic boundary conditions. 
During the structure optimization only lattice constants are modified, none of the  atoms change their position in the unit cell as all of them are placed in high symmetry crystallographic sites. 

\begin{table}[t]
\caption{The Strukturbericht symbols,  crystallographic space groups, and lattice parameters of the ordered Fe-Pt alloys with stoichiometric concentrations. The experimental values of lattice constants are presented in parenthesis. } 
\begin{tabular}{l|cccc}
             & Struktur- &  Space  & Lattice \\
             & bericht   &  Group  &  parameters\\
             & symbol    &  symbol &  (\AA)\\
\hline
FePt (FM)      & L1$_{0}$        & P4/mmm       &  $a=3.838$~(3.852) \\
               &                 &              &  $c=3.739$~(3.713) \\
Fe$_3$Pt (FM)  & L1$_{2}$        & Pm$\bar3$m   &  $a=3.737$~(3.750) \\
FePt$_3$ (FM)  & L1$_{2}$        & Pm$\bar3$m   &  $a=3.914$~(3.866) \\ 
FePt$_3$ (AFM) & L1$_{2}$        & Pm$\bar3$m   &  $a=3.891$~(3.866) \\
FePt$_3$ (NM)  & L1$_{2}$        & Pm$\bar3$m   &  $a=3.877$~(3.866) \\
\hline
\end{tabular}
\end{table}

The calculated lattice parameters are presented in Table I together with the Strukturbericht and space group symbols describing these structures.
The determined values differ slightly from experimental data placed in the parenthesis. The FePt and Fe$_3$Pt structures have been calculated assuming  ferromagnetic arrangement of the local magnetic moments. This is not the case for FePt$_3$ which is paramagnetic at room temperature. To investigate the influence of magnetic ordering on the lattice parameter of FePt$_3$, we have performed calculations for the FM, AFM, 
and nonmagnetic (NM) phases. Simple AFM ordering results in lattice parameter reduction, in comparison with the FM arrangement, however, the shortest lattice constant for FePt$_3$ structure has been obtained within the NM calculations.
 
\begin{table}[t]
\caption{The magnitude of the magnetic moments in $\mu_B$ calculated for three ordered Fe-Pt alloys with stoichiometric concentrations.} 
\begin{tabular}{l|cccc}
Atom  & Fe$_3$Pt (FM) & FePt (FM)  & FePt$_3$ (FM) & FePt$_3$ (AFM)\\
\hline
Fe & 2.711 & 2.966 & 3.258    & 3.325  \\
Pt & 0.349 & 0.354 & 0.359    & 0.000  \\
\hline
\end{tabular}
\end{table}

The magnetic moments calculated for the ordered Fe-Pt structures observed at room temperature are presented in Table II.
With increased iron concentration, magnetic moments on both types of atoms decrease and
the calculated values are in good agreement with the results presented previously. \cite{magn_FePt,mag-mom} 
The structure and magnetic ordering of the low-temperature phases are discussed in further sections.

\subsection{Phonon dispersion relations\label{subsec2}}

The phonon dispersion relations and PDOS were calculated with the direct method implemented in the PHONON code.\cite{Par97} 
This method utilizes the Hellmann-Feynman (H-F) forces obtained by performing small atomic displacements of nonequivalent atoms from their equilibrium positions. 
From them the dynamical matrix is determined and diagonalized to obtain the phonon frequencies at each wave vector. 
Since the DFT calculations are related to $T=0$~K, we have adopted the experimental lattice constants at room temperature
for comparison between the measured and calculated phonon frequencies. 

The experimental dispersion curves of the Fe-Pt alloys have been measured using  INS and published some decades ago. \cite{exp_FePt,exp_Fe3Pt,exp_FePt3}
The measurements were done for the relevant monocrystals with a high chemical order at room temperature. The phonon frequencies have been measured along the high symmetry
directions of the cubic symmetry: [1 0 0], [1 1 0] and [1 1 1], and additionally in [0 0 1] and [1 0 1] in the L1$_0$ tetragonal phase. In the L1$_2$ cubic phase, the latter two directions are equivalent to [1 0 0] and [1 1 0], respectively. For Fe$_3$Pt, the temperature and concentration dependence of the soft mode were also studied experimentally. \cite{kastner1, kastner2}

\begin{figure}[t]
\includegraphics[width=0.45\textwidth]{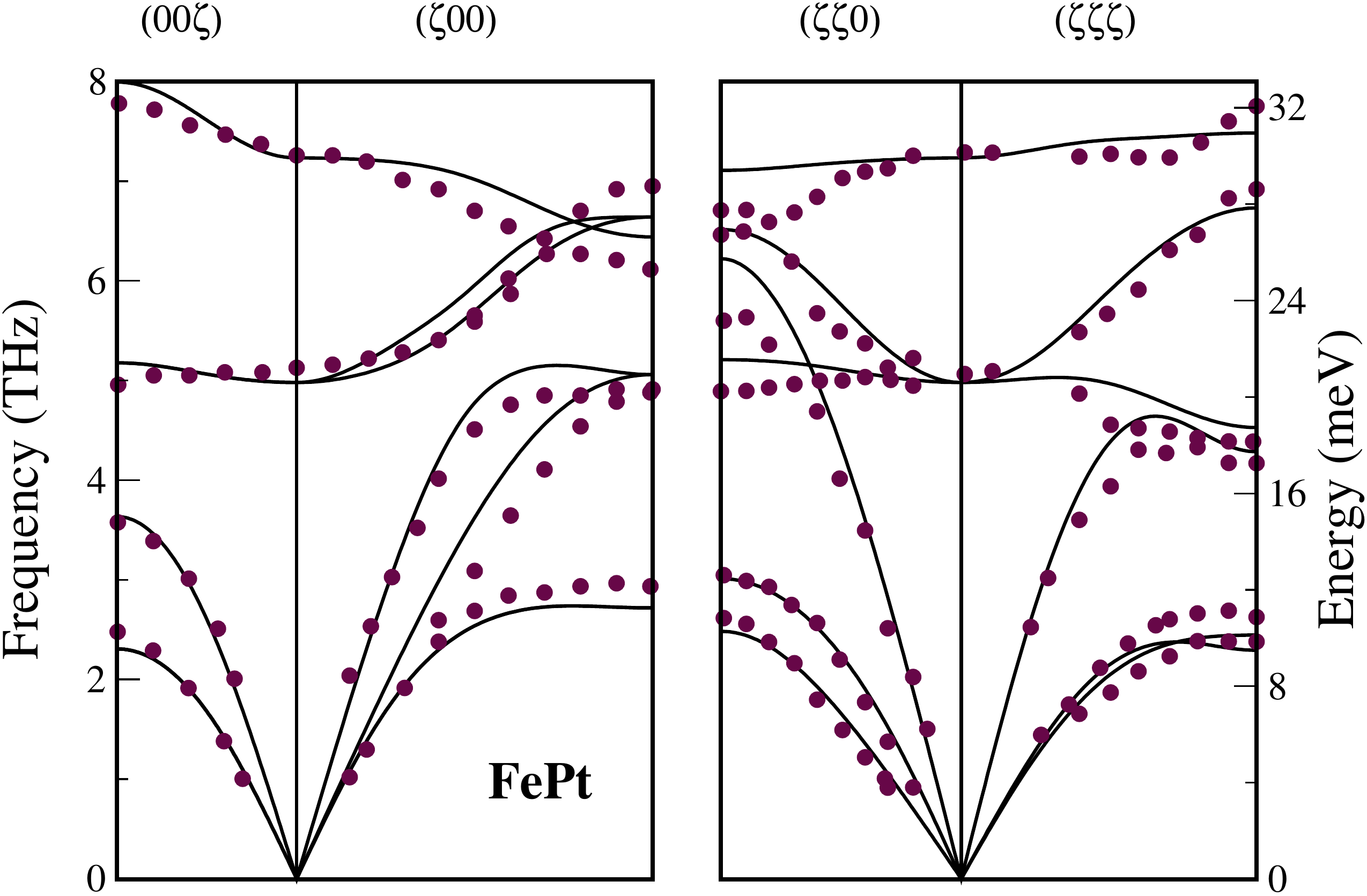}
\caption{\label{fig1_dysp_FePt}(Color online) The phonon dispersions along the high symmetry directions calculated for the L1$_0$ FePt structure. The experimental data are taken from Ref. [\onlinecite{exp_FePt}].}
\end{figure}

The conventional cell of the ordered L1$_0$ FePt structure contains four atoms, but it can be reduced to a primitive tetragonal cell with unchanged $c$ lattice constant and $a$ constant reduced to $a\sqrt2/2$. The primitive cell contains two atoms only, thus 6 branches of phonon dispersion curves are expected. They are presented in Fig.\ref{fig1_dysp_FePt} in comparison with the data measured by INS along the main crystallographic directions of the conventional cell (taken from Ref.~\onlinecite{exp_FePt}). The agreement between measured and calculated values is very good. The most significant discrepancies are observed for the high-frequency branches. For example, the calculated high-frequency dispersion curves along ($\xi$,$\xi$,0) and ($\xi$,$\xi$,$\xi$) directions are less dispersive than the related experimental curves. 

\begin{figure}[t]
\includegraphics[width=0.45\textwidth]{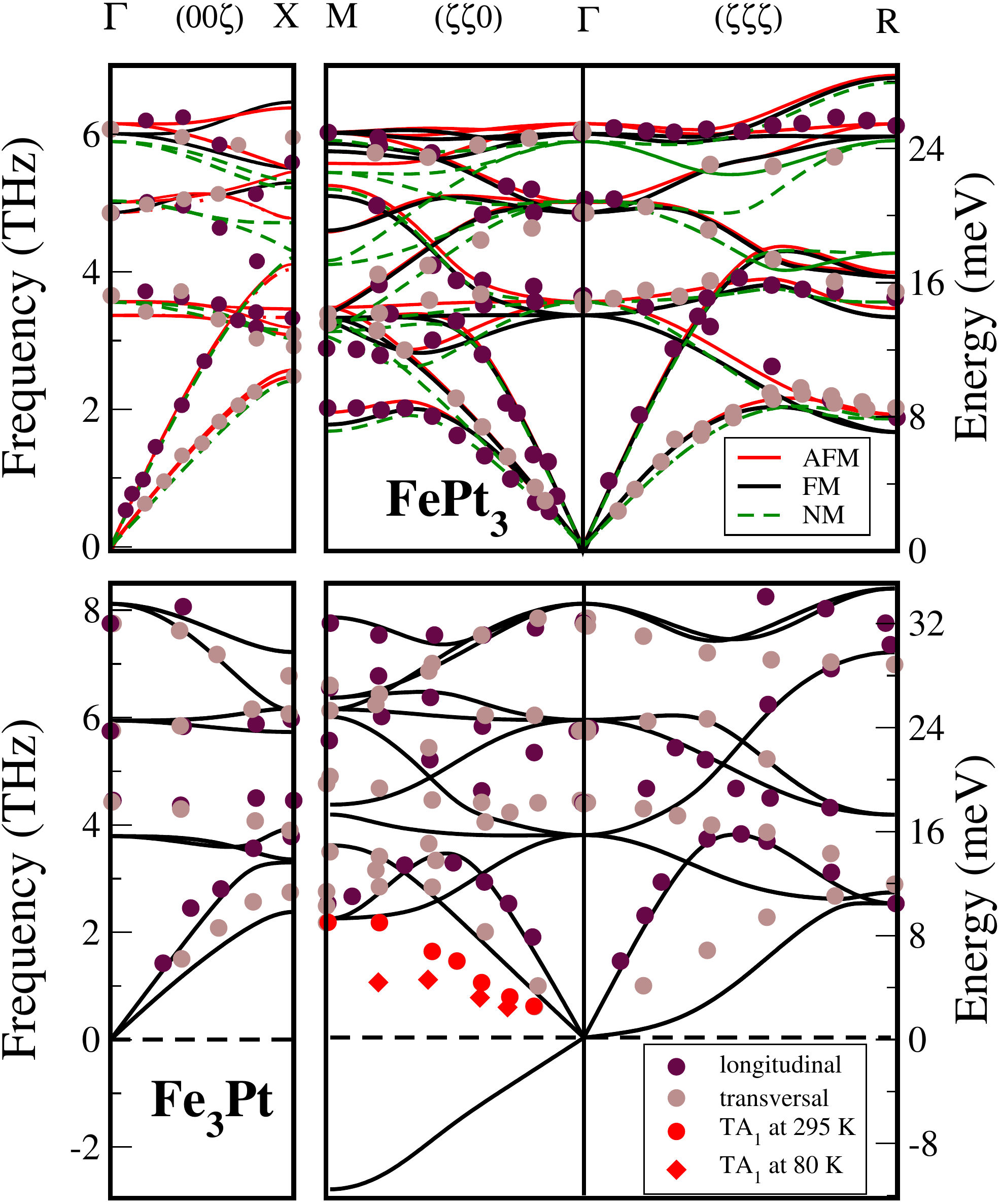}
\caption{\label{fig2_dysp_l12}(Color online) The phonon dispersions along the high symmetry directions calculated for the L1$_2$ structures of Fe$_3$Pt and FePt$_3$. 
The experimental data are taken from Refs. [\onlinecite{exp_FePt3}] and [\onlinecite{exp_Fe3Pt}].}
\end{figure}

The primitive cell of the L1$_2$ structure of Fe$_3$Pt or FePt$_3$ contains 4 atoms. Hence, one can expect 12 phonon dispersion curves. They are shown in Fig.~\ref{fig2_dysp_l12}. Different symbols of the experimental points correspond to two types of phonon polarizations: dark for longitudinal and light for transversal vibrations.  In the top panels, the experimental data taken for the FePt$_3$ alloy at 295 K \cite{exp_FePt3} are compared with dispersion relations calculated for three different magnetic phases: FM, AFM, and NM states. 
The frequencies of the AFM phase are slightly higher than those of the FM phase but differences between the dispersion curves are negligible.  Both of them reproduce the experimental data quite well, especially the frequencies calculated at the $\Gamma$ point are in very good agreement with measured values. The most significant discrepancies are observed for the highest optical curves  close to the $X$ and $R$ points. The NM calculations improve the agreement in the high-frequency region close to the $X$ point, but generally the NM state generates larger discrepancies than the FM or AFM states.  Each kind of calculations leads to too high frequencies at the $R$ point. Probably, the experimental signals were not registered there because of too low intensity in the Brillouin zone chosen in the experiment.

Moreover, in the measured curves some steep wiggles of the recorded frequencies of the longitudinal acoustic and the neighboring optical branches along the (0,0,$\xi$) direction have been noticed \cite{exp_FePt3}. For that acoustic mode, at the $X$ point, our calculations generate a frequency of 4~THz which is higher than 3.4 THz suggested from the fitting procedure of the Born-von Karman model.\cite{exp_FePt3} 
Additional analysis of the transverse acoustic branch along the ($\xi,\xi,\xi$) direction have been made as the downward bend of the curve measured close to the $R$ point could indicate the possible tetragonal instability at low temperatures.\cite{exp_FePt3} Our calculations do not show any anomalous behavior
at the $R$ point that could confirm this suggestion.

In the calculations performed for the Fe-rich L1$_2$ structure, the imaginary frequencies appear for the transversal acoustic mode along the ($\xi$,$\xi$,0) direction and approach the value 2.8i THz at the $M$ point (bottom panel of Fig.\ref{fig2_dysp_l12}). This behavior agrees with the measurements taken at 440, 295 and 80~K that showed significant frequency softening of the transversal acoustic (TA$_1$) mode registered along the ($\xi$,$\xi$,0) direction with decreasing temperature. \cite{exp_Fe3Pt} In the bottom panel of Fig.\ref{fig2_dysp_l12}, the dispersion curves measured at 295~K (dark and light points) are shown together with the TA$_1$ branch measured at 295~K and 80~K (red circles and diamonds, respectively).  The soft mode is described by the $M_2$ representation and is related to the vibration of Fe atoms only.\cite{exp_Fe3Pt} 
In Ref. \onlinecite{exp_Fe3Pt}, the authors noticed also the other mode ($M_4$) with the Fe atoms vibrating in a similar manner. The measured frequency of that optical mode is equal to the frequency of the $M_2$ mode at temperature of 440 K, and it stays almost constant with decreasing temperature. \cite{kastner1} In the present calculations, the $M_4$ mode vibrates with considerably higher frequency and this result confirms previous calculations. \cite{gruner} 

In calculations, the low-frequency acoustic branch along the ($\xi$,$\xi$,$\xi$) direction slightly softens in the vicinity of the $\Gamma$ point. 
The other acoustic branches and the high-frequency optical branches (between 6 and 8 THz) are well reproduced.  However, the position of the low-frequency optic branches (close to 4~THz) is intriguing. In this frequency range, all calculated optic branches are located significantly lower than the experimental data. Even at the $\Gamma$ point, almost 1 THz discrepancy is observed. Calculation errors can be excluded since the phonon dispersion relations are in very good agreement with the curves calculated previously by Gruner {\it et al.} \cite{gruner} The shift of phonon branches seems to be related to the martensitic phase transformation as well as the imaginary frequency at the $M$ point. The condensation of the soft mode drives the structure to a lower symmetry which is described in Sec.~\ref{Dis_subsec3}.

\subsection{Phonon density of states\label{subsec3}}

In previous studies, the phonon density of states has been deduced from the measured dispersion relations using the harmonic approximation and the Born-von Karman model.\cite{exp_FePt,exp_FePt3} 
The fitting procedure of the model to the measured data enables to determine the atomic force constants. Using the appropriate summations the total or partial (for different kinds of atoms) PDOS have been obtained for the L1$_0$ FePt structure. \cite{exp_FePt} The vibrational frequencies cover the region from 2 to 8 THz (8 to 32 meV) and as it can be expected the contribution of the heavier atoms to the total PDOS is dominant at low frequencies and that of the lighter atoms is important at high frequencies.

\begin{figure}[t]
\includegraphics[width=0.45\textwidth]{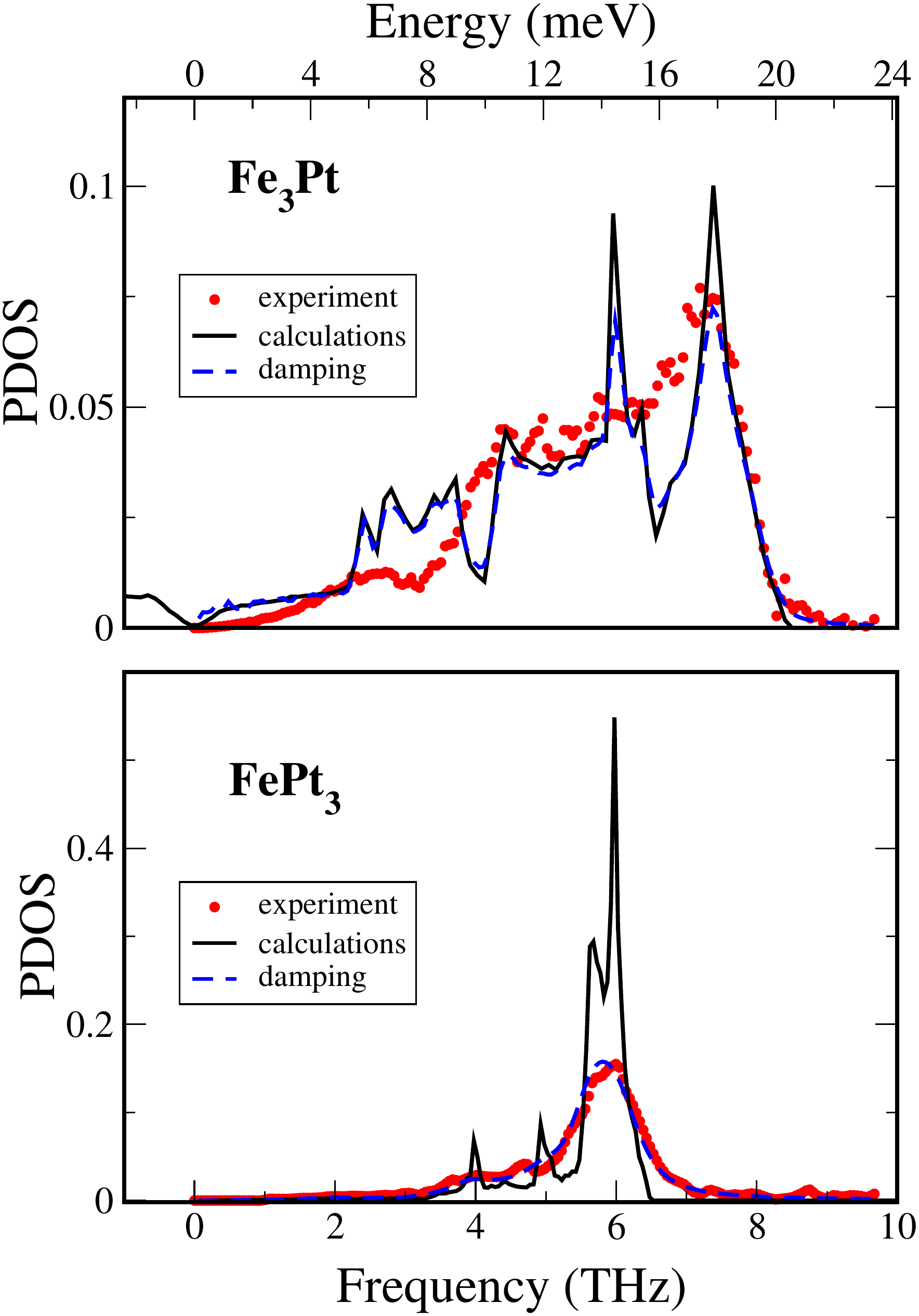}
\caption{\label{pdos_exp}(Color online) The Fe partial PDOS of Fe$_3$Pt and FePt$_3$ measured by NIS compared with the calculated spectra. The experimental data are taken from the NIS measurements performed as a part of this work.}
\end{figure}

The total PDOS of Fe$_3$Pt and FePt$_3$ have been shown and discussed in Ref. \onlinecite{exp_Fe3Pt}. The main difference between them is the high frequency limit of the spectra that is 6 and 8 THz for FePt$_3$ and Fe$_3$Pt, respectively. The partial spectra of those structures have not been calculated from the measured dispersion curves. 
 
The information on vibrational frequencies in Fe-Pt alloys has also been obtained from the NIS measurements. \cite{wiele,fultz,couet}  
For Fe$_3$Pt, the temperature dependent spectra were collected to study the low-temperature anomalies in that structure. \cite{wiele} For FePt$_3$, the NIS spectra of $^{57}$Fe atoms were measured on crystalline alloys that were chemically disordered, partially ordered, and L1$_2$ ordered. \cite{fultz} 
First experiments done on the bulk crystalline Fe$_3$Pt and FePt$_3$  samples provided the PDOS averaged over all crystallographic directions. \cite{wiele,fultz} The development of NIS technique accompanied with the layer by layer epitaxial preparation method allows to record the data even on monolayers of $^{57}$Fe atoms. \cite{Slezak,Stankov}  An appropriate choice of a substrate of the FePt thin films ordered in the L1$_0$ structure facilitated to perform the measurements of Fe partial PDOS along the (1,0,0) and (1,0,1) directions. \cite{couet} 
These experiments have shown the asymmetry of spectra in good agreement with the first principles calculations.

Currently, to complete the experimental data, the Fe partial PDOS of Fe$_3$Pt and FePt$_3$ have been measured by NIS at the P01 beamline of Petra III (Hamburg, Germany). The samples consisted of 30 nm films grown on MgO(100) substrates by co-evaporation of $^{57}$Fe (98\% isotopic enrichment) and Pt at a temperature of 500$^o$C. The measurements have been carried out at room temperature in grazing incidence geometry. The resolution of the monochromator that  defines the actual energy resolution of the measured data was 1 meV.   

In Fig. \ref{pdos_exp}, the measured Fe projected PDOSs are shown together with the data obtained from the calculations. In the case of the Fe$_3$Pt structure, the experimental phonon spectrum covers the wide frequency range from 2 up to 9 THz. The disagreement between the experimental and calculated data is mainly observed in the low-frequency region (2-4 THz), where the contribution of calculated PDOS is much larger than experimental one. 
This effect is related to the softening of modes calculated for the high-temperature austenite phase at 0~K. 
The high frequency end of the Fe$_3$Pt vibrational spectrum is well reproduced.
In the measured spectrum, a clearly visible shoulder at $\sim$2~THz corresponds to the TA$_1$ soft mode identified previously in the Fe$_3$Pt bulk sample \cite{exp_Fe3Pt} and nanoclusters. \cite{Pdos_Fe3Pt_nano}

In FePt$_3$, Fe atoms contribute to the PDOS in a much narrower range of frequencies with the main band centred around 6 THz, which is very well
reproduced in the calculations. Four sharp peaks noticeable in the theoretical spectrum are weakly visible in the experiment because of the broadening of each phonon line. The experimental resolution cannot explain such large broadening that is rather caused by the anharmonicity of a material or size and disorder effects in a measured specimen.  
The calculations are based on the harmonic approximation and performed for a perfect bulk crystalline sample. This implies an infinite phonon lifetime, which is not a likely situation to be found in a nanometric size sample of Fe-Pt alloys. The sample size, grain boundaries, and crystal defects reduce the lifetime. The effect of finite phonon lifetime can be approximated by applying a damped harmonic oscillator (DHO) model to the {\it ab initio} calculated phonon spectrum\,\cite{fultz}. This phenomenological model describes the "damping" of the phonons {\emph i.e.} it provides a finite phonon lifetime, which leads to broadening of phonon lines. 
Mathematically, this process is described as a convolution of the original PDOS with an appropriate broadening function $D(E,E*)$:
\begin{equation}\label{conv}
F_{DHO}(E*)=\int D(E,E*) F(E) \mathrm{d} E
\end{equation}
It has been shown that the damped harmonic oscillator function is a good choice for $D(E,E*)$\,\cite{fultz}:
\begin{equation}\label{eq34}
D(E,E*)= \frac{1}{\pi Q E*} \frac{1}{(\frac{E*}{E} - \frac{E}{E*})^2+\frac{1}{Q^2}}
\end{equation}
where $Q$ is the quality factor, which describes the number of periods a phonon will remain in a given state. Hence, the lower the value of $Q$, the shorter the lifetime and the larger broadening. This function is essentially a Lorentzian with a width that increases with $E*$, meaning that the PDOS features become broader at higher energy. The final PDOS (dashed blue line in Fig. \ref{pdos_exp}) is thus constructed by convolution of the theoretical PDOS with the DHO function using Eq.~(\ref{eq34}). 
In this model, only the $Q$ factor is adjusted to provide the best match between the theoretical and experimental PDOS. We fitted the calculation to the experimental data for Fe$_3$Pt using Q=30, which provides a better match of the high energy peak intensities. This rather large value of Q indicates that only little damping takes place in that particular thin film system.  
However, this broadening does not improve the agreement between the experiment and theory at lower frequencies that is caused by the soft-mode behavior. 
For FePt$_3$, a very good agreement is observed for Q=12, which indicates significant reduction of phonon lifetime in this system.

\section{Discussion\label{sec3}}

\subsection{Anisotropy of PDOS\label{Dis_subsec1}}

\begin{figure}[t]
\includegraphics[width=0.45\textwidth]{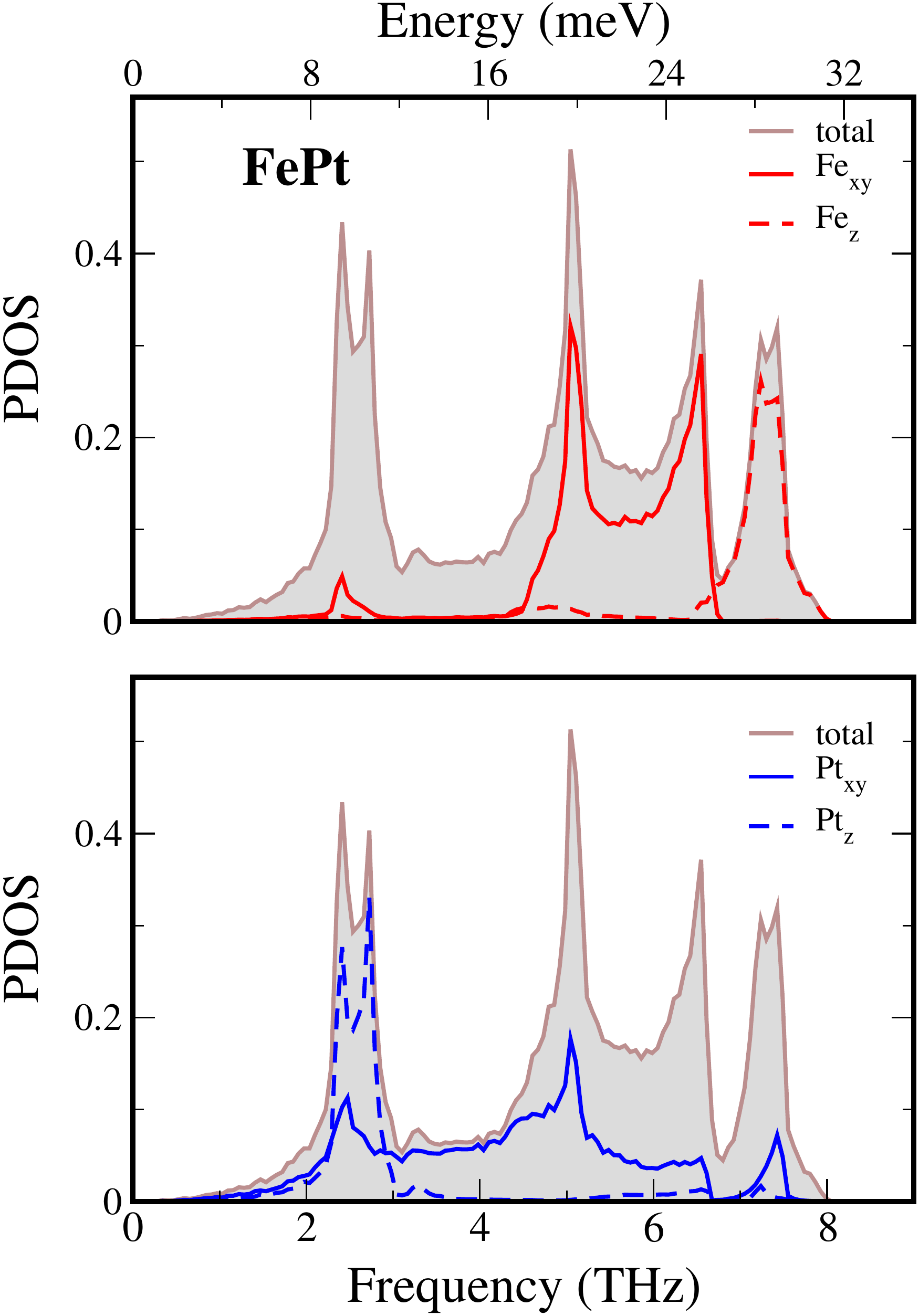}
\caption{\label{pdos_FePt}(Color online) The phonon density of states calculated for the L1$_0$ FePt structure. }
\end{figure}

\begin{figure*}[t]
\includegraphics[width=0.78\textwidth]{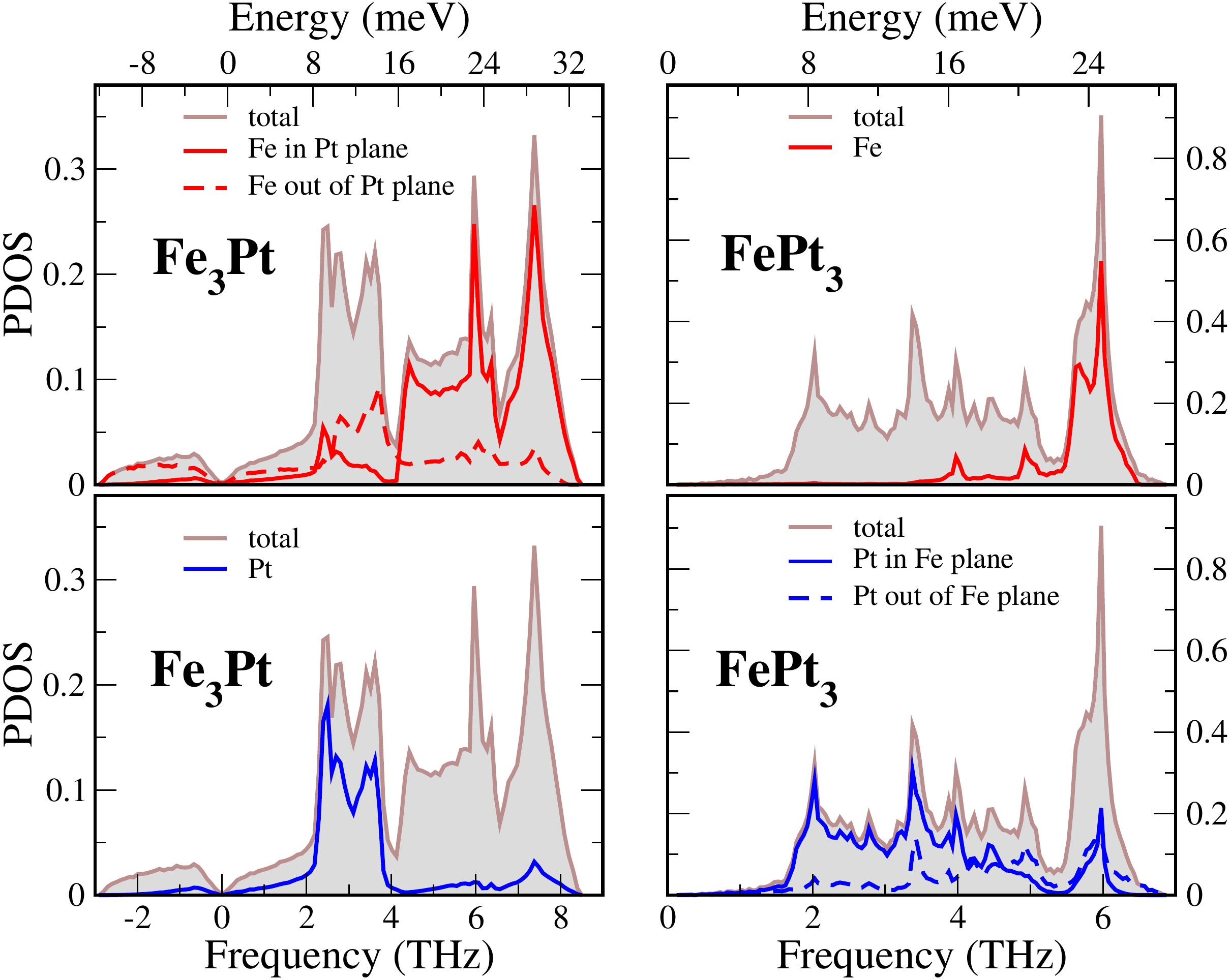}
\caption{\label{pdos_l12}(Color online) The phonon density of states calculated for the L1$_2$ structure of Fe$_3$Pt and FePt$_3$.}
\end{figure*}

Using first principles calculations one can determine the total PDOS as well as the  partial PDOS for different atoms and directions. In Figs. \ref{pdos_FePt} and \ref{pdos_l12}, the PDOS calculated for FePt, Fe$_3$Pt and FePt$_3$ are shown. Besides the total PDOS, the partial spectra of each type of atom projected on the three orthogonal directions [100], [010] and [001] are calculated. Due to atomic site symmetries some spectra are equivalent.  For the L1$_0$ tetragonal structure, the spectra taken along [100] and [010] are equivalent and their sum calculated separately for Fe or Pt atoms is shown in Fig.~\ref{pdos_FePt} as Fe$_{xy}$ or Pt$_{xy}$, respectively. Similarly, the spectra projected on the [001] direction are named Fe$_z$ or Pt$_z$. The strong asymmetry of Fe vibrations shown in Fig.~\ref{pdos_FePt} has already been observed in the NIS experiment.\cite{couet} The vibrations along two directions are separated almost completely and the vibrations in the {\it z} direction contribute to a high frequency peak around 7.2~THz. 
In contrast, the Pt atoms vibrations along the {\it z} axis dominate in a narrow frequency range from 2 to 3 THz, while those in the {\it xy}-plane are spread in the whole spectrum including the peak at the lowest frequency.

For the L1$_2$ structures, there are two kinds of atomic sites, atoms placed at corners with cubic site symmetry and atoms situated in the center of cubic faces with tetragonal symmetry. For Pt (or Fe) atoms placed in cubic symmetry sites of the Fe$_3$Pt (or FePt$_3$) unit cell the PDOS projected on three crystallographic directions are equivalent. For the remaining three Pt (or Fe) atoms placed in the center of a cubic face, two equivalent directions lying in each unit face are discriminated as "in Pt (or Fe) plane" and the third direction, perpendicular to the unit face is called "out of Pt (or Fe) plane" (Fig.\ref{pdos_l12}). The negative values of frequencies shown in the left panels of Fig.\ref{pdos_l12} are related to the soft mode existing in Fe$_3$Pt. The soft mode dominantly consists of Fe vibrations in the "out of Pt plane" direction. The Pt atom vibrations are mainly observed in the narrow frequency range between 2 and 4 THz. There is a significant difference between the Fe partial PDOS collected along the "in Pt plane" and "out of Pt plane" directions.  The "out of plane" vibrations are involved in the low frequency region and the "in plane" vibrations are distributed at higher frequencies.

The spectrum calculated for Fe atoms in FePt$_3$ is characterized by a high frequency peak at about 6 THz with high intensity and two significantly smaller peaks at 4 and 5 THz. The pronounced differences are observed for the PDOS calculated for Pt atoms vibrating along the direction parallel ("in Fe plane") or perpendicular to the cell face ("out of Fe plane").  

Comparing the spectra for the three Fe concentrations one can see that the highest frequency of the total PDOS decreases with decreasing Fe content.

\subsection{Low-temperature AFM phases of FePt$_3$\label{Dis_subsec2}}

\begin{table*}[t]
\caption{The space group, nonequivalent atomic positions, magnitude of magnetic moments, and energy per 4 atoms unit cell calculated for the ordered FePt$_3$ alloy with different magnetic orders.} 
\begin{ruledtabular}
\begin{tabular}{lccccc}
magnetic&space 	& lattice 	& non-equivalent	& magnetic 	& energy per\\
order	& group	& parameters (\AA)& atoms		& moments ($\mu_B$)& unit cell (eV)\\
\hline
FM   & $Pm\bar{3}m$ & $a=3.914$ & Fe(0,0,0)    & 3.258 & 0\\
     &       &         & Pt($\frac{1}{2}$,$\frac{1}{2}$,0)    & 0.359 &  \\
AFM  & $Pm\bar{3}m$ & $a=3.891$ & Fe(0,0,0)    & 3.325 & 0\\
     &       &         & Pt($\frac{1}{2}$,$\frac{1}{2}$,0)    & 0.006 &     \\
NM  & $Pm\bar{3}m$ & $a=3.887$ & Fe(0,0,0)    & 0.0 & 1.101\\
     &       &         & Pt($\frac{1}{2}$,$\frac{1}{2}$,0)    & 0.0 &     \\
AFM-I & $P4/mmm$ & $a=5.533$ & Fe(0,0,0)    & 3.272 & -0.094\\
     &       & (3.912) & Fe($\frac{1}{2}$,$\frac{1}{2}$,0)    & 3.272  &  \\
     &       & $c=3.900$        & Pt($\frac{1}{4}$,$\frac{1}{4}$,$\frac{1}{2}$)    & 0.000 &  \\
AFM-II & $P4/mmm$ & $a=3.926$ & Fe(0,0,0)    & 3.306 &-0.081\\
     &       & $c=7.756$ & Fe(0,0,$\frac{1}{2}$)    & 3.306 &  \\
     &       & (3.878) & Pt(0,$\frac{1}{2}$,$\frac{1}{4}$)    & 0.000 &  \\
     &       &         & Pt($\frac{1}{2}$,$\frac{1}{2}$,0)    & 0.000 &  \\
     &       &         & Pt($\frac{1}{2}$,$\frac{1}{2}$,$\frac{1}{2}$)    & -0.080 &  \\
\hline
\end{tabular}
\end{ruledtabular}
\end{table*}

\begin{figure}[h!]
\includegraphics[width=0.45\textwidth]{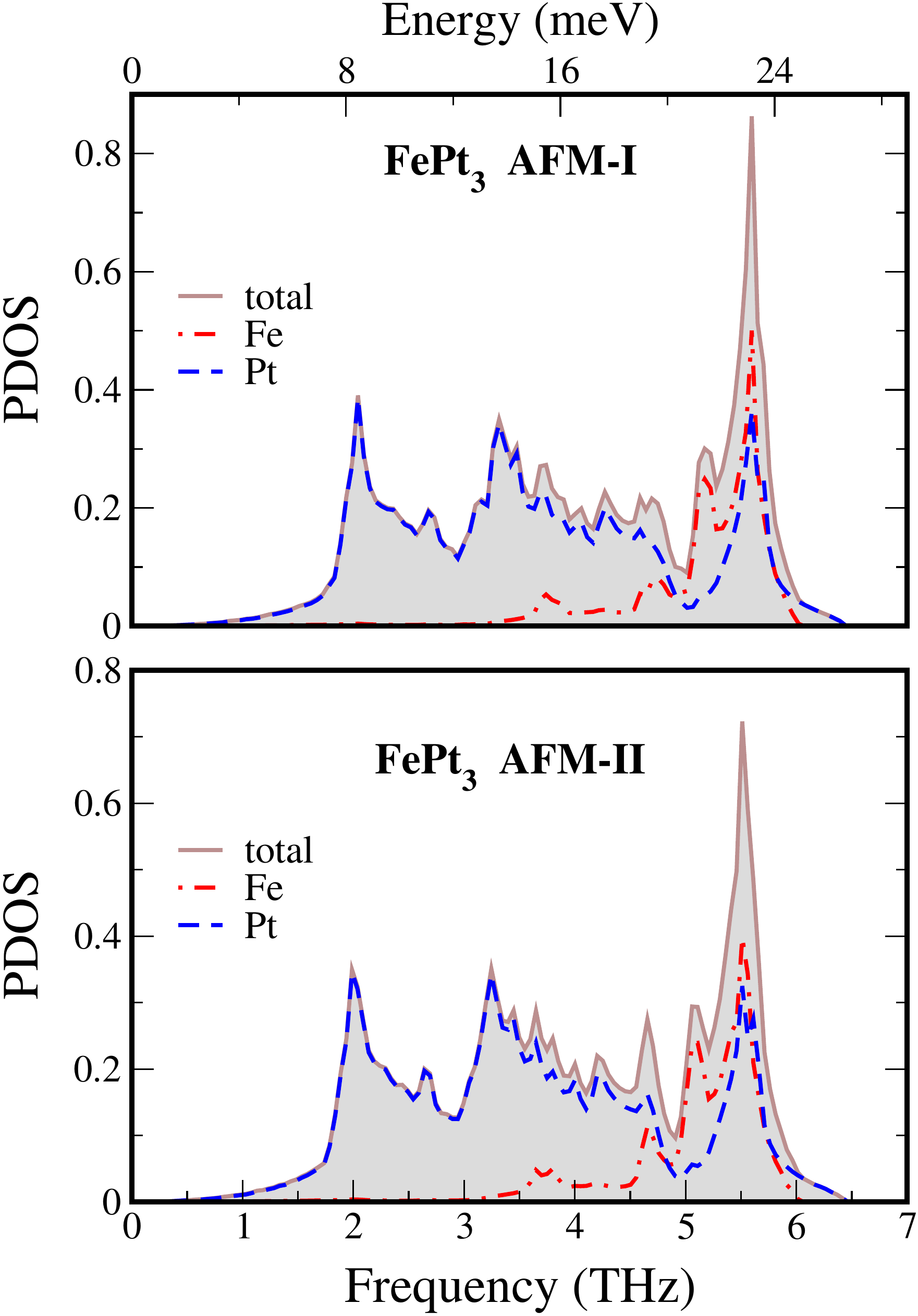}
\caption{\label{pdos_AFM}(Color online) The phonon density of state calculated for the FePt$_3$ low-temperature phases.}
\end{figure}

The chemically disordered phase of the Fe-Pt alloy with stoichiometry 1:3 is ferromagnetic with Fe moments of 2~$\mu_B$. 
With decreasing temperature, the alloy transforms from the disordered fcc to ordered L1$_2$ structure with paramagnetic properties. 
Further decreasing the temperature leads to the arrangement of magnetic moments within two different types of AFM order. 
In the bulk, the appearance of these two AFM phases depends strongly on the deviation from the ideal composition.\cite{AFM_1,AFM_3} 
The FePt$_3$ phase is maintained in Fe$_x$Pt$_{1-x}$ alloys for 0.22$\leq$x$\leq$0.41. \cite{str_FePt3} 
Upon cooling, the AFM-I order develops below 160 K, where the Fe moments order in alternating ferromagnetic (110) planes. This transition is of second order. The AFM-I tetragonal unit cell is obtained by doubling the cubic unit cell with the basic vectors $a_t$=(1,1,0)$a_c$, $b_t$=(1,-1,0)$a_c$ and $c_t$=(0,0,1)$a_c$.   
The moments carried by the Fe atoms are equal to 3.3~$\mu_B$. 
With decreasing temperature, this phase transforms subsequently to the AFM-II phase where the ferromagnetically ordered (001) planes are stacked antiferromagnetically. 
It results in doubled lattice constant in the c-direction and the orientation of tetragonal unit cell is $a_t$=(1,0,0)$a_c$, $b_t$=(0,1,0)$a_c$  and $c_t$= (0,0,2)$a_c$. 
The symmetry of both phases is characterized by the P4/mmm space group, however, the magnetic unit cells and positions of non-equivalent atoms are different (Table III). 

In Table III, the structural data, magnetic moments and calculated energy of the AFM-I and AFM-II supercells are compared with the artificial FM, AFM, and NM structures of FePt$_3$. 
The calculations performed for both AFM phases show that the ground state energy of the AFM-I phase is slightly lower than that of AFM-II phase, however, the energy difference between AFM-I and AFM-II is not appreciable. Similarly, the energies of artificial high-temperature structures of FM and AFM magnetic order are almost the same. The magnitudes of magnetic moments of Fe atoms vary from 3.258 $\mu_B$ to 3.325 $\mu_B$ for the cubic phases and from 3.272 $\mu_B$ to 3.306 $\mu_B$ for tetragonal AFM structures. As it is shown in Table III, the lattice parameters of FePt$_3$ in AFM-I stay almost unchanged.  In parenthesis, the values related to the cubic lattice constant are presented to make comparison easier. In the AFM structure, these values are smaller by about 0.05-0.1\% than the cubic lattice constant. In the AFM-II structure, the $a$ lattice constant elongates by 0.3\% and $c/2$ shortens by 0.9\% in comparison to the cubic structure. 

Since the interatomic distances and the magnetic moments stay almost unchanged, also the interatomic forces and the PDOS presented in Fig.~\ref{pdos_AFM} change very weakly (apart from small differences observed around 4.7 THz). Both of them are similar to the spectrum calculated for the FM phase and the shape of the partial Fe PDOS is also the same. The small shift to lower frequencies is caused by lattice constants used in the calculations of AFM-I and AFM-II, which differ slightly from experimental parameters applied for the cubic structure.   

In the present studies, no stresses in the structure are assumed, in contradiction with a previous publication where authors speculate about the existence of AFM-I and AFM-II ordering under the uniaxial and tetragonal stresses. \cite{magnet_FePt$_3$}  The central issue of that first principle calculations was to investigate the influence of strain on the magnetic state.

\subsection{Low-temperature phase of Fe$_3$Pt\label{Dis_subsec3}}

The structure of the low-temperature phase of stoichiometric Fe$_3$Pt ordered alloy has not yet been exactly described.  
X-ray diffraction investigations performed by different groups on polycrystals with similar composition and degree of atomic order so far did not yield conclusive results on the low temperature structure. \cite{kastner2} This material cannot be investigated by means of neutron diffraction since Fe and Pt have almost the same coherent scattering lengths, 9.45 fm and 9.60 fm, respectively.
Up to now, three types of martensitic phases have been reported:  bcc, bct, and fct.\cite{kastner1,yamamoto1,yamamoto} The bcc martensite is formed when the degree of order of the parent phase is very low. If the degree of order is in the intermediate range, a thermoelastic transformation from the parent phase occurs and the bct martensite is formed. For an ordering degree of 0.80, the fct martensite forms through a second order transformation. The latest X-ray diffraction measurements revealed that the highly ordered Fe$_3$Pt alloy exhibits a martensitic transformation below 60 K and the tetragonality parameter $c/a$ of the low temperature phase is larger (1.005) or lower (0.94) than 1 for the degree of order 0.88 or 0.75, respectively.  \cite{yamamoto}

In the present calculations, the deformation caused by uniaxial stress leading to the tetragonal fct ($P4/mmm$) phase with c/a larger and lower than 1 has been considered. The optimization of those structures allowed to find structures with energy slightly lower than that for the cubic phase (see Table IV). The optimization procedure finished when the magnitudes of the H-F forces were less than 10$^{-5}$ eV/\AA\ and the external pressure reached zero. After optimization,  small stresses are observed in both tetragonal structures: 3.83 kb in [100] and [010] directions and -6.94 kb in [001] direction for tetragonality $c/a>$1 and -0.95 kb and 1.91 kb for $c/a<$1. The tetragonal deformation slightly modifies the magnitude of magnetic moments of Fe atoms, mainly those lying in the (110) plane. They increase with increasing lattice constant $c$. This kind of tetragonal deformation cannot remove the imaginary frequencies (2.552i THz for cubic structure) from dispersion curves at the $M$ point, it lays even deeper at 2.806i THz. 

\begin{figure}[b]
\includegraphics[width=0.45\textwidth]{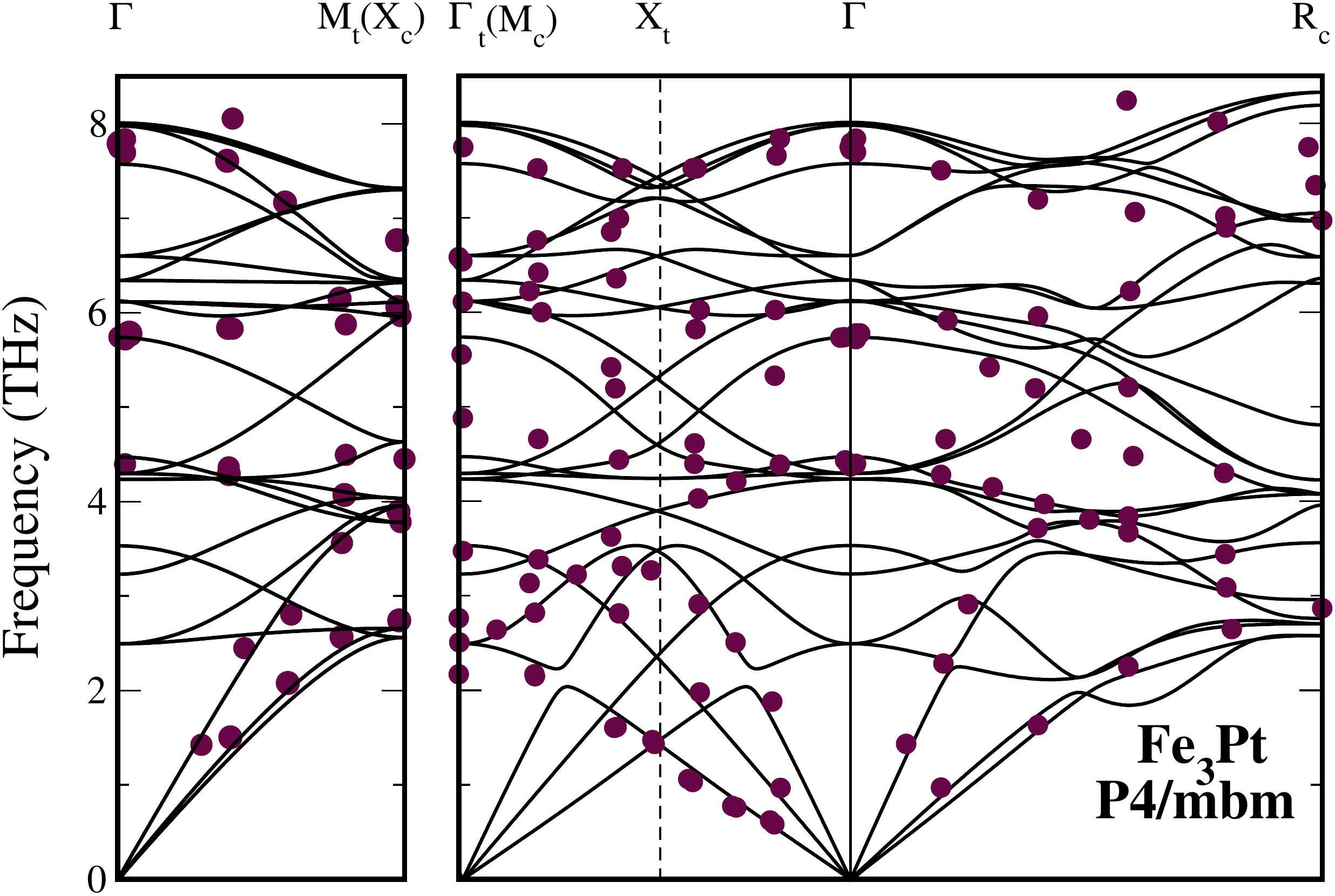}
\caption{\label{fonon_127}(Color online) The phonon dispersion curves calculated for tetragonal structure of Fe$_3$Pt. The experimental data are taken from Ref. [\onlinecite{exp_Fe3Pt}].}
\end{figure}

To stabilize the crystal structure, we have distorted the cubic lattice using the soft-mode polarization vector and then optimized
the lattice parameters with new symmetry constraints. Previously, the calculations with the distorted Fe$_3$Pt crystal have been performed to investigate the changes in electronic structure induced by the soft-mode \cite{gruner}.
The resulting crystal structure has a lower total energy and reduced symmetry to the tetragonal $P4/mbm$ space group (No 127) with a doubled unit cell. 
The dispersion curves calculated using this space group are presented in Fig.~\ref{fonon_127}. The high symmetry directions of the cubic phase (Fig.\ref{fig2_dysp_l12}) are used since the geometries of the cubic and tetragonal phases are comparable. The number of curves is 24 as there are 8 atoms in the primitive cell. The experimental data measured at room temperature for cubic Fe$_3$Pt are compatible with our calculated dispersion curves. Especially, the TA$_1$ mode which is soft in the cubic phase is perfectly reproduced by the lowest acoustic branch calculated for the tetragonal phase. Additionally, the measured optic modes with too high frequencies in the cubic structure can be pinned to some curves obtained from these calculations. 

\begin{table*}[t]
\caption{The comparison of high and low-temperature phases of completely ordered Fe$_3$Pt.} 
\begin{ruledtabular}
\begin{tabular}{lccccc}
magnetic&space 	& lattice 	& non-equivalent	& magnetic 	& energy per\\
order	& group	& parameters (\AA)& atoms		& moments ($\mu_B$)& unit cell (eV)\\
\hline
above M$_s$  & $Pm\bar3m$ & $a=3.737$ & Fe(0,0,0)    			    & 2.710-2.714 & 0\\
             &       &         & Pt($\frac{1}{2}$,$\frac{1}{2}$,0)    & 0.349 &  \\
\hline
below M$_s$     & $P4/mbm$ & $a=5.253$ & Pt(0,0,0)    			    & 0.346 & -0.019\\
             &        & (3.910) & Fe(0,$\frac{1}{2}$,0) 	    & 2.704 & \\ 
             &        & $c=3.789$ & Fe(0.2275,0.2725,0) 		    & 2.721-2.728& \\
          & $P4/mmm$  & $a=3.701$ & Fe($\frac{1}{2}$,0,$\frac{1}{2}$)   & 2.711 & -0.008\\
          &        & $c=3.813$ & Fe(0,$\frac{1}{2}$,$\frac{1}{2}$)    & 2.710 & \\ 
          &        &         & Fe($\frac{1}{2}$,$\frac{1}{2}$,0)    & 2.734 &     \\
          &        &         & Pt(0,0,0) 		    	    & 0.357& \\
          & $P4/mmm$  & $a=3.776$ & Fe($\frac{1}{2}$,0,0)    	    & 2.729 & -0.016\\
          &        & $c=3.664$ & Fe(0,$\frac{1}{2}$,$\frac{1}{2}$)    & 2.729 & \\ 
          &        &         & Fe($\frac{1}{2}$,$\frac{1}{2}$,0)    & 2.669   \\
          &        &         & Pt(0,0,0) 		    	    & 0.352& \\
\hline
\end{tabular}
\end{ruledtabular}
\end{table*}

\begin{figure}[t]
\includegraphics[width=0.45\textwidth]{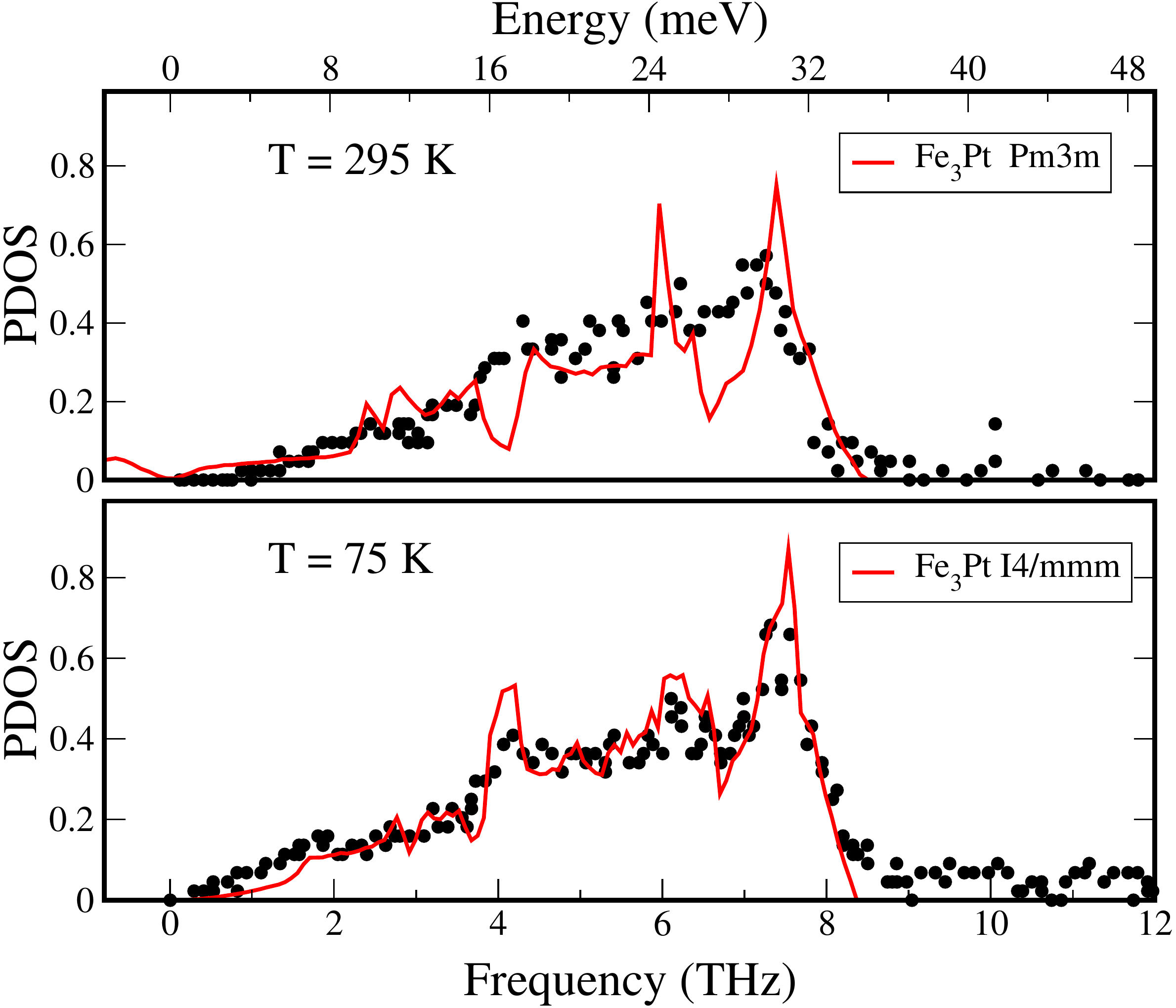}
\caption{\label{pdos_127}(Color online) The phonon density of states calculated for the tetragonal $P4/mbm$ structure of Fe$_3$Pt.
The experimental points are taken from Ref. [\onlinecite{wiele}].}
\end{figure}

The anomalous temperature dependence of the PDOS in the Fe$_3$Pt ordered alloys has been studied using the NIS technique.\cite{wiele} The presence of the soft mode leads to enhancement of the intensities of the low-frequency vibrations so in the low-frequency region the PDOS proceeds to zero not in a parabolic but linear manner. In Fig. \ref{pdos_127}, the Fe PDOS measured by Wiele {\it et al.} \cite{wiele}
at $T=295$ and 75~K are presented together with the spectra calculated for the cubic $Pm\bar{3}m$ and tetragonal $P4/mbm$ structures, respectively.
The agreement between the theory and experiment is much better for the tetragonal phase measured at 75~K than for the cubic one at room temperature.
The positions and intensities of three peaks observed in the experiment correspond very well to those found
theoretically. Also, the shoulder at about 2 THz originating from the soft mode is very well reproduced in calculations.
It clearly demonstrates that the tetragonal structure derived from the soft-mode is an important constituent of the low-temperature phase.  
It may be related to static precursors of the martensitic phase observed by X-ray, electron, and neutron diffraction 
and manifested as tetragonally strained structures growing in the austenite phase approaching the structural transformation \cite{muto,foos,seto}. 

It should be noted that this type of crystal deformation may be driven by the electron-phonon interaction.  
According to previous theoretical studies on Fe$_3$Pt, the condensation of the soft-mode opens a pseudogap in the density of states
and leads to the extended reconstruction of the Fermi surface due to nesting effects \cite{gruner}.
Such electron-phonon coupling involving the soft-mode distortion and changes in valence electron density 
may be a characteristic feature of other Invar alloys, which exhibit martensitic transformation at low temperatures.

\section{Conclusions\label{sec4}}
 
To complete and systematize the understanding of the ordered Fe$_3$Pt, FePt, and FePt$_3$ alloys, first-principles studies of their structural, magnetic, and dynamical properties have been performed. The obtained results have been compared with existing experimental data supplemented by the new partial Fe PDOS measured in thin Fe-Pt films of L1$_2$ structure using the nuclear inelastic scattering technique. The calculations involve the room temperature as well as the low temperature phases of Fe-Pt alloys. 

Conclusions about the room temperature structures of ordered Fe-Pt alloys can be summed up as follows:
 
(i) The phonon dispersion relations calculated for FePt and FePt$_3$ show very good agreement with inelastic neutron scattering data measured at room temperature. The paramagnetic FePt$_3$ systems were modelled using static FM and AFM arrangements of local magnetic moments on Fe atoms  
and both approaches reproduce the experimental data much better than the calculations for the non-magnetic state. In the cubic L1$_2$ structure of Fe$_3$Pt, the observed TA$_1$ soft mode at the $M$ point of the Brillouin zone leads to some discrepancy between the calculated and measured phonon branches. 

(ii) On the basis of the analysis of different variants of calculated partial PDOSs, the experimentally unobservable anisotropy of some sites or atoms is noticed.  The anisotropy of the partial Fe and Pt PDOSs appears evidently in the tetragonal FePt, however, it is also observed for the atomic sites of the tetragonal symmetry in the cubic Fe$_3$Pt and FePt$_3$. 
For these cubic structures the PDOS asymmetry, which is not important for a bulk material, can influence the properties of nanocrystals or surfaces. 

(iii) The Fe projected PDOS calculated for the Fe$_3$Pt and FePt$_3$ cubic structures were compared with new nuclear inelastic scattering measurements for thin films at room temperature. For FePt$_3$, we found very good agreement between experiment and theory although additional broadening of the measured PDOS is observed. 
The soft-mode behavior observed in Fe$_3$Pt generates some disagreement between the theoretical and experimental PDOS, mainly in the low-frequency region.

The chemical ordering in conventional bulk Fe-Pt alloys becomes very slow at low temperatures making the experimental studies of the low temperature structures very difficult. Nevertheless, this region was extensively investigated for the Fe-rich compositions, as these alloys show the Invar effect and undergo a martensitic transformation, and for the Pt-rich compositions, where the paramagnetic-antiferromagnetic phase transition is observed. 
The results of our research on the low-temperature phases of Fe$_3$Pt and FePt$_3$ can be summarized as follows:

(i) In FePt$_3$, the crystal structure parameters, magnetic moments, and total energies were determined for 
two antiferromagnetic arrangements that induce the tetragonal distortion ($P4/mmm$). The ground state energies of AFM-I and AFM-II configurations are nearly the same and slightly lower than the energies of the hypothetical FM and AFM phases. The magnitude of magnetic moments calculated on Fe atoms is about 3.3 $\mu_B$ for AFM-I and AFM-II. In contrast, the measured magnetic moment of AFM-II phase is 2~$\mu_B$ and is equal to the value in disordered Fe-Pt alloys. It proves the previous findings describing the AFM-II phase observed in FePt$_3$ as formed by antiphase domains. \cite{maat_AFM_FePt$_3$} Our calculations showed that any  antiferromagnetic structure is energetically favoured and the changes induced by these two magnetic configurations affect very weakly the lattice dynamical properties leading to very similar phonon spectra. 

(ii) In Fe$_3$Pt, the low-symmetry tetragonal structure ($P4/mbm$) has been derived using the soft-mode polarization vector.
The energy of this structure is lower than the energies of two tetragonal fct phases ($P4/mmm$) obtained by uniaxial stress deformation. A few meV per unit cell differences between them seems to be irrelevant for the stabilization of the low temperature phase, however, 
it was shown that the dispersion curves and the Fe projected PDOS calculated for the $P4/mbm$ symmetry agree very well with
the inelastic neutron scattering and nuclear inelastic scattering (at $T=75$~K) data, respectively. 
It shows that although the martensitic phase transition is not a group-subgroup transformation, the tetragonal structure derived
from the soft mode plays an important role in this transition as the main constituent of the low-temperature phase of Fe$_3$Pt. This is also the case in the martensitic transformation of NiTi alloys where the soft mode of the austenite phase leads to either an intermediate incommensurate phase locked into a
trigonal R phase or to an orthorhombic phase, which in turn, creates a low-frequency mode and favours the monoclinic martensitic phase. \cite{NiTi}

\begin{acknowledgments}

The authors acknowledge support by the COST Action MP0903 
"Nanoalloys as Advanced Materials: From Structure to Properties and Applications" 
and by the Polish National Science Center (NCN) under Project No. 2011/01/M/ST3/00738. We also acknowledge financial support from the Found for Scientific Research - Flouders (FWO) and the Concerted Research Action (GOA/14/007) at KU Leuven.

\end{acknowledgments}


\begin{thebibliography}{99}

\bibitem{ph_diag} T. B. Massalski (editor), Binary Alloy Phase Diagrams (ASM International, Ohio 1990) p. 1752.

\bibitem{invar} E. F. Wassermann, in {\it Ferromagnetic Materials},
edited by K. H. Buschow and E. P. Wohlfarth (Elsevier, Amsterdam, 1990) Vol. 5, p. 237.  

\bibitem{martensite} S. Kajiwara and W. S. Owen, Metal. Trans. \textbf{5}, 2047 (1974). 

\bibitem{tajima} K. Tajima, Y. Endoh, Y. Ishikawa and W. G. Stirling, 
			\prl \textbf{37}, 519 (1976).

\bibitem{maat_AFM_FePt$_3$} S. Maat, O. Hellwig, G. Zeltzer, E. E. Fullerton, G. J. Mankey, M. L. Crow and J. L. Robertson,
		 \prb \textbf{63}, 134426 (2001).
	

\bibitem{exp_FePt} V. Pierron-Bohnes, R. V. P Montsouka, C. Goyhenex, T.Mehaddene, L. Messad, H. Bouzar, H. Numakura, K. Tanaka, and B. Hennion, Defect and Diffusion Forum \textbf{1}, 263 (2007).

\bibitem{exp_FePt3} Y. Noda, Y. Endoh, S. Katano, and M. Izumi,
                   Physica B+C \textbf{120}, 317 (1983). 

\bibitem{exp_Fe3Pt} Y. Noda and Y. Endoh, 
                   J. Phys. Soc. Jpn.  \textbf{57}, 4225 (1988).

\bibitem{kastner1} J. Kastner, W. Petry, S. M. Shapiro, A. Zheludev, J. Neuhaus, Th. Roessel, E. F. Wassermann, and H. Bach,
		Eur. Phys. J. B \textbf{10}, 641 (1999).

\bibitem{kastner2} J. Kastner, J. Neuhaus, E. F. Wassermann, W. Petry, B. Hennion, and H. Bach,
		Eur. Phys. J. B \textbf{11}, 75 (1999).


\bibitem{wiele} N. Wiele, H. Franz , W. Petry,
		Physica B {\bf 263}, 716 (1999)

\bibitem{fultz} B. Fultz, T. A. Stephens, E. E. Alp, M. Y. Hu, J. P. Sutter, T. S. Toellner, and W. Sturhahn,
		\prb {\bf 61}, 14517 (2000)
		
\bibitem{Pdos_Fe3Pt_nano} B. Roldan Cuenya, J. R. Croy, L. K. Ono, A. Naitabdi, H. Heinrich, W. Keune, J. Zhao, W. Sturhahn,
E. E. Alp, and M. Hu, \prb {\bf 80}, 125412 (2009),
                                     
\bibitem{tamada} Y. Tamada, R. Masuda, A. Togo, S. Yamamoto, Y. Yoda, I. Tanaka, M. Seto, S. Nasu, and T. Ono, Phys. Rev. B {\bf 81}, 132302 (2010).     
                                
\bibitem{couet} S. Couet, M. Sternik, B. Laenens, A. Siegel, K. Parlinski, N. Planckaert, F. Grostlinger, A. I. Chumakov, R. R\"{u}ffer, B. Sepiol, K. Temst, and A. Vantomme, 
		\prb \textbf{82}, 094109 (2010).

\bibitem{ptj} P. T. Jochym, K. Parlinski, and A. M. Ole\'{s}, Eur. Phys. J. B \textbf{61}, 173 (2008).

\bibitem{anisotropy_FePt} M. Tsujikawa, A. Hosokawa, and T. Oda,
                 \prb \textbf{77}, 054413 (2008).

\bibitem{magnetism_FePt} J. Honolka, T. Y. Lee, K. Kuhnke, A. Enders, R. Skomski, S. Bornemann, S. Mankovsky, J. Minar, J. Staunton,H. Ebert, M. Hessler, K. Fauth, G. Schutz, A. Buchsbaum, M. Schmid, P. Varga, and K. Kern,    \prl \textbf{102}, 067207 (2009).

\bibitem{gruner} M. E. Gruner, W. A. Adeagbo, A. T. Zayak, A. Hucht, and P. Entel, \prb \textbf{81}, 064109 (2010).		 
		 
\bibitem{vasp}  G. Kresse and J. Furthm\"uller,
                   Comput. Mater. Sci. \textbf{6}, 15 (1996);
                   \prb \textbf{54}, 11169 (1996).

\bibitem{Per96} J. P. Perdew, K. Burke, and M. Ernzerhof, \prl \textbf{77}, 3865 (1996).

\bibitem{magn_FePt} Zhihong Lu, R. V. Chepulskii, and W. H. Butler,
		\prb \textbf{81}, 094437 (2010).

\bibitem{mag-mom} V. N. Antonov, B. N. Harmon, and A. N. Yaresko,
		\prb \textbf{64}, 024402 (2001).                                       

\bibitem{Par97} K. Parlinski, Z. Q. Li, and Y. Kawazoe,
                    \prl \textbf{78}, 4063 (1997);
                 K. Parlinski,
                   Computer code {\sc phonon}, Cracow, 2012.

\bibitem{Slezak} T. \'{S}lezak, J. \L{}a\.{z}ewski, S. Stankov, K. Parlinski, R. Reitinger, M. Rennhofer, R. R\"{u}ffer, 
B. Sepiol, M. \'{S}lezak, N. Spiridis, M. Zajac, A.I. Chumakov, and J. Korecki, Phys. Rev. Lett. {\bf 99}, 066103 (2007).

\bibitem{Stankov} S. Stankov, R. Rohlsberger, T. \'{S}lezak, M. Sladecek, B. Sepiol, G. Vogl, A.I. Chumakov, R. R\"{u}ffer, 
N. Spiridis, J. \L{}a\.{z}ewski, K. Parlinski, and J. Korecki, Phys. Rev. Lett. {\bf 99}, 185501 (2007).
                   
                                                                
\bibitem{AFM_1} G. E. Bacon and J. Crangle, Proc. R. Soc. London. Ser.
A. 272, 387 (1963).

\bibitem{AFM_3} D. Palaith, C. W. Kimball, R. S. Preston, and J. Crangle, Phys.
Rev. {\bf 178}, 795 (1969).

\bibitem{str_FePt3} O. Kubschewski, Iron-Binary Phase Diagrams Springer, New York, (1982).

\bibitem{magnet_FePt$_3$} Dongyoo Kim and Jisang Hong, Journal of Magnetics {\bf 16}, 197 (2011).

\bibitem{yamamoto1} M. Yamamoto, T. Fukuda, T. Kakeshita, K. Koyama, H. Nojiri, 
		Phys. Procedia \textbf{10} 117 (2010).
                
\bibitem{yamamoto} M. Yamamoto, S. Sekida, T. Fukuda, T. Kakeshita, K. Takahashi,
K. Koyama, Hi. Nojiri, J. Alloys and Compounds \textbf{509}, 8530 (2011).        
                   
\bibitem{muto} S. Muto, R. Oshima, F.-E. Fujita,  Met. Trans.  {\bf 19 A}, 2727 (1988).
                    
\bibitem{foos} M. Foos, C. Frantz, and M. Gantois, Scripta Metall.  {\bf 12}, 795 (1978).
                   
\bibitem{seto} H. Seto, Y. Noda, and Y. Yamada, J. Phys. Soc. Japan  {\bf 59}, 965, 978 (1990).

\bibitem{NiTi} K. Parlinski and M. Parlinska-Wojtan, \prb \textbf{66}, 064307 (2002).


                   
\end{thebibliography}
\end{document}